\documentclass[A4,10pt]{iopart}
\usepackage{amsfonts}
\usepackage{amssymb}
\usepackage{geometry}
\usepackage{graphicx}
\usepackage{dsfont}

\begin{document}

\title{Truncated $\gamma$-exponential models for tidal stellar systems}
\author{Y. J. Gomez-Leyton and L. Velazquez}
\address{Departamento de F\'\i sica, Universidad Cat\'olica del Norte, Av. Angamos 0610, Antofagasta,
Chile.}

\begin{abstract}
We introduce a parametric family of models to characterize the properties of astrophysical systems in a quasi-stationary evolution under the incidence evaporation. We start from an one-particle distribution $f_{\gamma}\left(\mathbf{q},\mathbf{p}|\beta,\varepsilon_{s}\right)$ that considers an appropriate deformation of Maxwell-Boltzmann form with inverse temperature $\beta$, in particular, a power-law truncation at the scape energy $\varepsilon_{s}$ with exponent $\gamma>0$. This deformation is implemented using a generalized $\gamma$-exponential function obtained from the \emph{fractional integration} of ordinary exponential. As shown in this work, this proposal generalizes models of tidal stellar systems that predict particles distributions with \emph{isothermal cores and polytropic haloes}, e.g.: Michie-King models. We perform the analysis of thermodynamic features of these models and their associated distribution profiles. A nontrivial consequence of this study is that profiles with isothermal cores and polytropic haloes are only obtained for low energies whenever deformation parameter $\gamma<\gamma_{c}\simeq 2.13$.
E-mail: yuvineza.gomez@gmail.com; lvelazquez@ucn.cl
\end{abstract}

\section{Introduction}

Gravitation is an example of attractive interaction that is unable to confine particles. It is always possible that some particles acquire sufficiently energy (via encounters or any other mechanism) and scape from gravitational influence of the system. Evaporation is an unavoidable process that drives dynamical evolution of astrophysical systems\footnote{By itself, a closed system of particles interacting via gravitation or Van del Waals forces undergo the incidence of evaporation. The key question here is the existence of a \emph{finite potential barrier} to scape from that system. For a system in fully equilibrium, a Maxwell distribution of velocities with finite temperature always exhibits a fraction of particles that overcomes any finite potential barrier. Evaporation can be avoided introducing an external field with an infinite potential barrier, as example, a container with impenetrable walls. While the presence of a container with impenetrable walls is very usual consideration in physical laboratories to confine substances that are found in a gaseous phase, its is nonphysical argument in real astrophysical situations. Therefore, astrophysical systems only can reach a quasi-stationary evolution because of the incidence of evaporation.}. As discussed elsewhere \cite{Binney,Spitzer,michie,king,chandra,Vel.QEM1,Vel.QEM2,wooley}, the incidence of evaporation is a very important ingredient to explain the behavior of stellar structures observed in Nature. The effects of evaporation crucially depend on the relaxation mechanisms that are present in the microscopic dynamics, which depend on concrete conditions of a given particular system. Among such mechanisms, one can mention (i) the particles collisions \cite{michie,king,chandra}, which are present in stellar systems sufficiently dense such as globular clusters; and (ii) general mechanisms like \emph{parametric resonance}, which is the origin of chaotic behavior of microscopic dynamics in nonlinear many-body Hamiltonian systems with bound motions in the configuration space \cite{Cipriani}.

We propose in this work a parametric family of astrophysical models that accounts for the influence of evaporation under different relaxation regimes: the \emph{truncated $\gamma$-exponential models}. These phenomenological models constitutes a suitable generalization of some models of tidal stellar systems available in the literature, such as Wooley \& Dickens truncated isothermal model \cite{wooley}, as well as King's models of globular clusters \cite{king}. This proposal enable us to analyze the influence of the evaporation on the thermodynamic behavior and distribution profiles. Moreover, it provides a unification framework of the known isothermal and polytropic profiles considered in hydrodynamic models of stellar systems. Remarkably, these models are sufficiently amenable to allow an analytical derivation of most of their associated hydrodynamic quantities.

The paper is organized into sections as follows. Second section is devoted to introduce the parametric family of quasi-stationary distributions, the derivation of their associated hydrodynamic quantities, as well as some details about the integration of the associated nonlinear Poisson equation. Afterwards, third section is devoted to discuss numerical results obtained from this model, both information concerning to thermodynamical description as well as distribution profiles. Some conclusions are drawn in forth section. Finally, mathematical properties concerning to the so-called $\gamma$-\emph{exponential function} are discussed in \ref{Exp.Append}.

\section{A parametric family of models}

Both observational evidences and theoretical analysis suggest that distribution profiles of many stellar systems can be explained considering a quasi-stationary one-particle distribution with a mathematical form close to Maxwell-Boltzmann distribution \cite{Hjorth}:
\begin{equation}\label{MB.prof}
f_{MB}\left(\mathbf{q},\mathbf{p}\right)  \sim\exp\left[-\beta\varepsilon(\mathbf{q},\mathbf{p})\right],
\end{equation}
being $\varepsilon(\mathbf{q},\mathbf{p})=\mathbf{p}^{2}/2m+m\phi\left(  \mathbf{q}\right)$ the mechanical energy of a given particle with momentum $\mathbf{p}$ located at the position $\mathbf{q}$ with a mean field gravitational potential $\phi\left(  \mathbf{q}\right)$. The incidence of evaporation introduces a deformation of the above distribution, specifically, a truncation for energies above a certain \emph{scape energy} $\varepsilon_{s}$. Two examples of distributions that introduce this kind of deformation are the truncated isothermal model proposed by Wooley and Dickens \cite{wooley}:
\begin{equation}\label{WD.prof}
f_{WD}\left(\mathbf{q},\mathbf{p}|\beta,\varepsilon_{s}\right)  =
A\exp\left\{  \beta\left[  \varepsilon_{s}-\varepsilon(\mathbf{q},\mathbf{p})\right] \right\},
\end{equation}
as well as Michie-King models \cite{michie,king}:
\begin{equation}\label{MK.prof}
f_{MK}(\mathbf{q},\mathbf{p}|\beta,\varepsilon_{s})=
A\left\{  \exp\left\{  \beta\left[ \varepsilon_{s}-\varepsilon(\mathbf{q},\mathbf{p})\right]
\right\}  -1\right\},
\end{equation}
which vanish for energies $\varepsilon(\mathbf{q},\mathbf{p})>\varepsilon_{s}$, where $A$ represents a normalization constant. The first one corresponds to a Maxwell-Boltzmann distribution (\ref{MB.prof}) that is discontinuously truncated at the scape energy $\varepsilon_{s}$, while the second one considers a progressive vanishing at this point.

\begin{figure}
  \centering
  \includegraphics[width=4.5in]{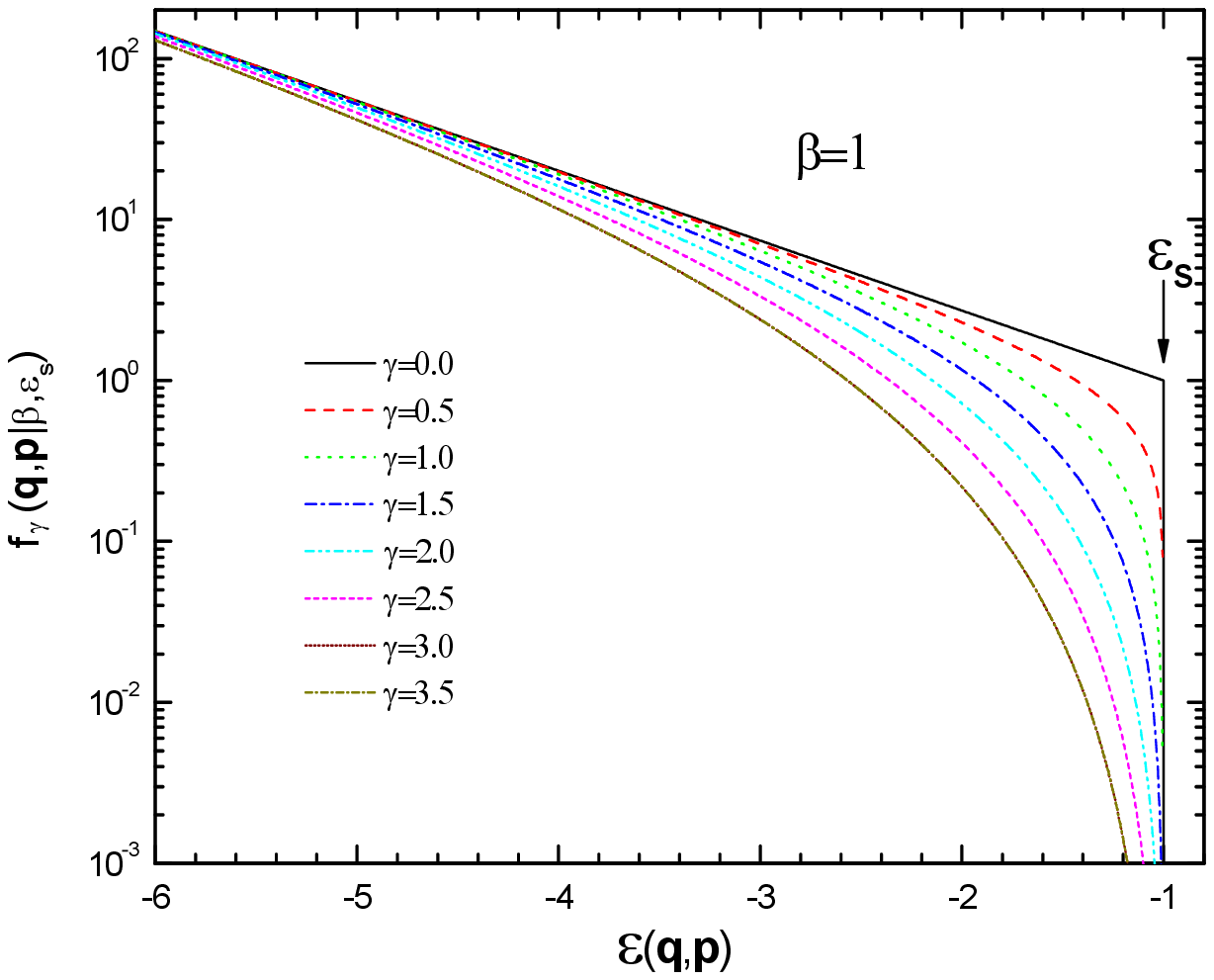}\\
  \caption{One-particle distributions \emph{versus} mechanical energy $\varepsilon(\mathbf{q},\mathbf{p})$ corresponding to truncated $\gamma$-exponential model (\ref{df.model}) for some values of deformation parameter $\gamma$. We have employed prefixed values $\varepsilon_{s}=-1$ and $\beta=1$ for the scape energy and the inverse temperature, respectively.}\label{gamma.distribution.eps}
\end{figure}

Our interest is to propose a generalization of evaporation truncation scheme considered in the above astrophysical models. Firstly, let us express them into a unifying form as follows:
\begin{equation}\label{df.model}
f_{\gamma}(\mathbf{q},\mathbf{p}|\beta,\varepsilon_{s})=A E_{\gamma}(x),
\end{equation}
where $x=\beta\left[ \varepsilon_{s}-\varepsilon(\mathbf{q},\mathbf{p})\right]$, while $E_{\gamma}(x)$ is a certain function obtained from the \emph{truncation of exponential function} $\exp(x)$, with $\gamma$ being a certain parameter. In particular, distributions (\ref{WD.prof}) and (\ref{MK.prof}) are described by the functions:
\begin{equation}\label{f01}
E_{0}(x)=\left\{
           \begin{array}{cc}
             \exp(x), & \mbox{for }x>0, \\
             0, & \mbox{otherwise}, \\
           \end{array}
         \right.\mbox{ and }
E_{1}(x)=\left\{
           \begin{array}{cc}
             \exp(x)-1, & \mbox{for }x>0, \\
             0, & \mbox{otherwise}. \\
           \end{array}
         \right.
\end{equation}
These two truncation schemes can be generalized for any nonnegative integer $\gamma=n$ subtracting the first $n$ terms of power-expansion of exponential function:
\begin{equation}\label{n.truncated}
E_{n}(x)=\left\{
           \begin{array}{cc}
             \exp(x)-\sum^{n-1}_{k=0}x^{k}/k!, & \mbox{for }x>0, \\
             0, & \mbox{otherwise}. \\
           \end{array}
         \right.
\end{equation}
Noteworthy that this ansazt contains the functions (\ref{f01}) as particular cases with $\gamma=0$ and $\gamma=1$, which now guarantees the vanishing of its $n-1$ first derivatives at $x=0$:
\begin{equation}\label{first.derivatives}
E_{n}(0)=E_{n}'(0)=\ldots=E_{n}^{(n-1)}(0)=0.
\end{equation}
For positive values of the argument $x$, the function (\ref{n.truncated}) can be rewritten into a more convenient integral representation as follows:
\begin{equation}
E_{n}(x)=\frac{1}{(n-1)!}\int^{x}_{0}\eta^{n-1}\exp(x-\eta)d\eta.
\end{equation}
This integral representation enables a direct extension of the above truncation scheme for any nonnegative real number $\gamma$ replacing the factorial $(n-1)!$ by the known Gamma function $\Gamma(\gamma)$:
\begin{equation}
\Gamma(\gamma)=\int^{+\infty}_{0}\eta^{\gamma-1}\exp(-\eta)d\eta.
\end{equation}
The resulting function:
\begin{equation}\label{integral.gamma}
E_{\gamma}(x)\equiv E\left(  x;\gamma\right)=\frac{1}{\Gamma(\gamma)}\int^{x}_{0}\eta^{\gamma-1}\exp(x-\eta)d\eta= \sum_{n=0}^{\infty}\frac{x^{\gamma+n}}{\Gamma\left(  \gamma+1+n\right)}
\end{equation}
modifies the usual exponential function using $\gamma$ as a \emph{deformation parameter}. Hereinafter, this function will be referred to as \emph{$\gamma$-exponential function}, and the same one will be equivalently denoted as $E_{\gamma}(x)$ or $E\left( x;\gamma\right)$. Definition (\ref{integral.gamma}) is simply the \emph{Riemann-Liouville fractional integral} of the ordinary exponential function \cite{Miller}. Considering the fractional differentiation rule for the power-function series:
\begin{equation}
\frac{d^{\alpha}}{dx^{\alpha}}\sum_{\mu}a_{\mu}x^{\mu}\equiv\sum_{\mu}a_{\mu}\frac{\Gamma(\mu+1)}{\Gamma(\mu-\alpha+1)}x^{\mu-\alpha},
\end{equation}
conditions (\ref{first.derivatives}) are now generalized as follows:
\begin{equation}
\frac{d^{\alpha}}{dx^{\alpha}}E\left(0;\gamma\right)=0,
\end{equation}
where order of fractional differentiation $\alpha$ belongs to the interval $0\leq \alpha<\gamma$. Readers can find some useful properties of $\gamma$-exponential function $E\left(  x;\gamma\right)$ in \ref{Exp.Append}. 

The proposed family of quasi-stationary distributions (\ref{df.model}) describe the influence of evaporation with two independent parameters: the scape energy $\varepsilon_{s}$ and the deformation parameter $\gamma$. This family of distributions is schematically shown in figure \ref{gamma.distribution.eps}. The same ones exhibit a Maxwell-Boltzmann profile (\ref{MB.prof}) for energies $\varepsilon(\mathbf{q},\mathbf{p})$ that are far enough from below the scape energy $\varepsilon_{s}$. Besides, they show a \emph{power-law truncation} at $\varepsilon_{s}$ with exponent $\gamma$. The increasing of deformation parameter $\gamma$ characterizes a larger deviation from Maxwell-Boltzmann profile (\ref{MB.prof}). For a phenomenological viewpoint, a larger value of the deformation parameter $\gamma$ describes a stronger influence of evaporation, or equivalently, the incidence of a weaker relaxation mechanism. As shown below, the ansatz (\ref{df.model}) can describe distribution profiles with isothermal cores and polytropic haloes, where deformation parameter $\gamma$ \emph{determines the polytropic structure of haloes}. The scape energy $\varepsilon_{s}=m\phi_{s}$ can be derived from the called \emph{tidal potential} $\phi_{s}$, which accounts for the gravitational influence of neighboring systems \cite{Binney,Spitzer,michie,chandra}. According to this energetic truncation, a particle that is trapped under the system gravitational influence will be confined inside a finite region of the space, where gravitational potential $\phi(\mathbf{q})$ fulfills the inequality $\phi(\mathbf{q})\leq\phi_{s}$. For a system with spherical symmetry, the boundary of this region is the sphere of radius $R_{t}$:
\begin{equation}\label{tidal radius}
R_{t}=-GM/\phi_{s},
\end{equation}
which is referred to as the \emph{tidal radius} \cite{Binney}, with $M=Nm$ being the total mass of astrophysical system.

Let us now obtain some hydrodynamic observables associated with family of distributions (\ref{df.model}). Introducing the dimensionless variable $\tau=\beta p^{2}/2m$ as well as the dimensionless potential $\Phi=\Phi\left(  \mathbf{q}\right)$:
\begin{equation}\label{dimensionless.pot}
\Phi\left(  \mathbf{q}\right)=\beta m\left[\phi_{s}-\phi\left(  \mathbf{q}\right)\right],
\end{equation}
the particles density $n(\mathbf{q})$ can be calculated as follows:
\begin{equation}
n\left(  \mathbf{q}\right)  =\int f_{\gamma}\left(
\mathbf{q},\mathbf{p}\right)
d^{3}\mathbf{p}=A^{\prime}\int_{0}^{\Phi}E\left(
\Phi-\tau;\gamma\right) \tau^{\frac{1}{2}}d\tau,
\end{equation}
being $A^{\prime}=2\pi A\left(  2m/\beta\right)  ^{\frac{3}{2}}$. The integration can be performed by using the convolution formula (\ref{convolution}) with parameter $\nu=3/2$, which leads to express the particles density in terms of the $\gamma$-exponential function as follows:
\begin{equation}\label{densidad particulas}
n\left(  \mathbf{q}\right)  =n\left[\Phi(\mathbf{q});\gamma\right]  =CE\left[
\Phi(\mathbf{q});\gamma+\frac{3}{2}\right].
\end{equation}
Here, $C=A\left(  2m\pi/\beta\right)^{\frac{3}{2}}$ and $\Gamma(3/2)=\sqrt{\pi}/2$, in accordance with gamma function for half-integer values:
\begin{equation}
\Gamma\left(n + \frac{1}{2}\right) = \frac{(2n)!}{4^{n} n!}\sqrt{\pi}.
\end{equation}
Other important observable is the kinetic energy density $\upsilon(\mathbf{q})$:
\begin{equation}
\upsilon(\mathbf{q})=\int f_{\gamma}\left(  \mathbf{q},\mathbf{p}\right)  \frac{1}{2m}%
\mathbf{p}^{2}d^{3}\mathbf{p}=\frac{A^{\prime}}{\beta}\int_{0}^{\Phi}%
E\left(  \Phi-\tau;\gamma\right)  \tau^{\frac{3}{2}}d\tau,
\end{equation}
which is also expressed in terms of the $\gamma$-exponential function using the convolution formula (\ref{convolution}) with parameter $\nu=5/2$:
\begin{equation}
\upsilon(\mathbf{q})=\upsilon\left[\Phi(\mathbf{q});\gamma\right]=\frac{3}{2\beta}CE\left[  \Phi(\mathbf{q});\gamma+\frac{5}{2}\right].
\end{equation}
Finally, we can introduce the magnitude $\epsilon (\mathbf{q})=\upsilon(\mathbf{q})/n(\mathbf{q})$:
\begin{equation}
\epsilon(\mathbf{q}) = \frac{3}{2\beta}\chi\left(\Phi;\gamma\right)
=\frac{3}{2\beta}\frac{E\left(\Phi;\gamma+\frac{5}{2}\right)}
{E\left(\Phi;\gamma+\frac{3}{2}\right)},
\end{equation}
which represents the kinetic energy per particle.

\begin{figure}[t]
\begin{center}
\includegraphics[width=4.5in]{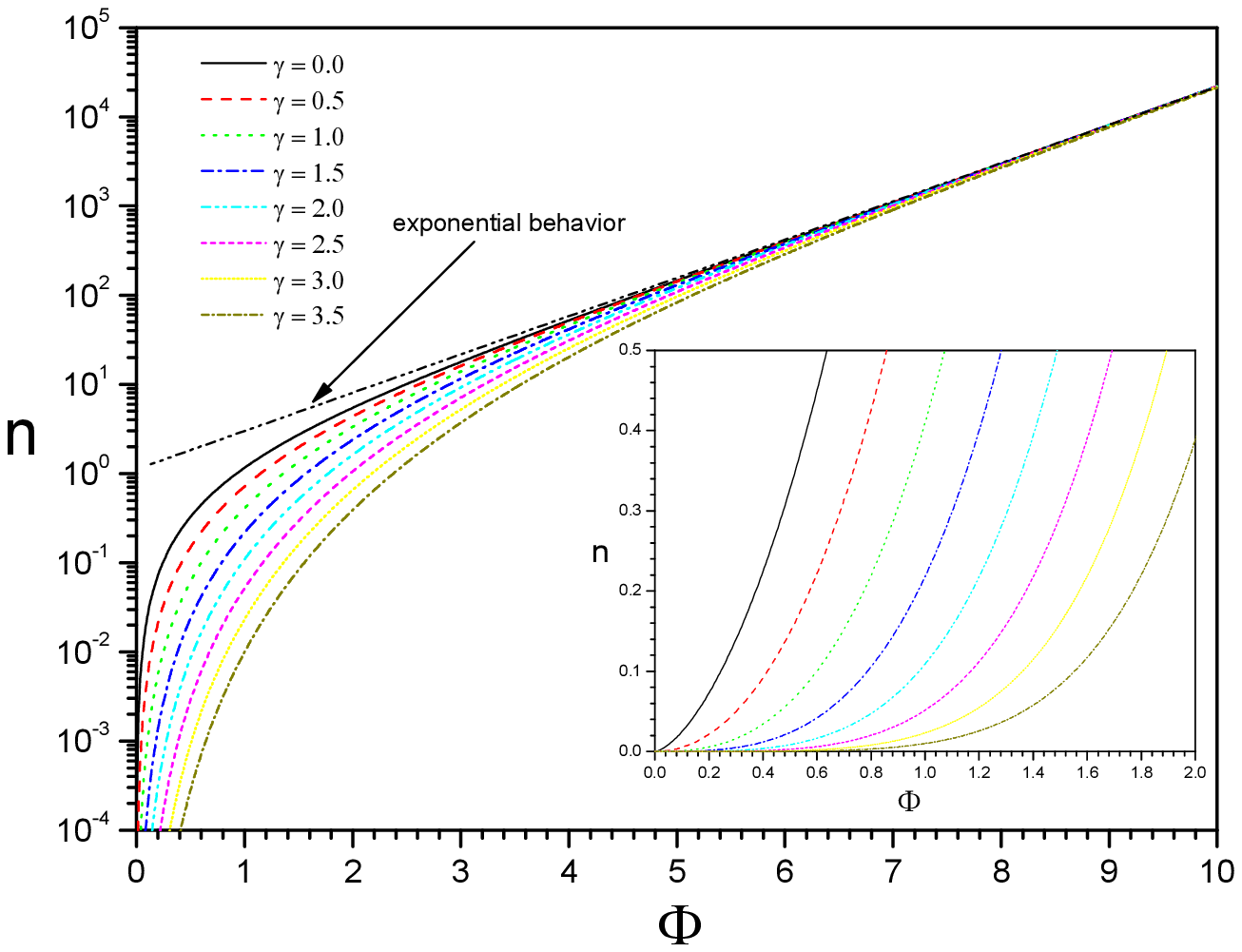}
\caption{Dependence of particles density $n(\Phi,\gamma)$ on the dimensionless potential $\Phi$  for some values of deformation parameter $\gamma$. One can notice the asymptotic convergence of all these dependencies towards an exponential growth for large values of $\Phi$, $n(\Phi,\gamma)\propto\exp(\Phi)$, which is related to a dominance of isothermal conditions. We show in the inset panel these same dependencies in order to show the power-law behavior for small values of $\Phi$, $n(\Phi)\propto\Phi^{\gamma}$, which is related to a dominance of polytropic conditions as a consequence of evaporation. This type of dependence can describe distributions with isothermal cores and polytropic haloes.}
\label{densidad.eps}
\end{center}
\end{figure}

According to the kinetic theory for ideal gases, this magnitude $\epsilon(\mathbf{q})$ can be related with the fluid pressure $p(\mathbf{q})$ as:
\begin{equation}\label{PresionGamma}
p(\mathbf{q}) = \frac{2}{3}n(\mathbf{q})\epsilon
(\mathbf{q})=\frac{1}{\beta}CE\left(  \Phi;\gamma+\frac{5}{2}\right).
\end{equation}
The hydrostatic equilibrium takes place in a fluid when gravitational force and pressure gradient are balanced as follows:
\begin{equation}\label{hidro}
\nabla p(\mathbf{q})= \rho(\mathbf{q}) \textbf{g}(\mathbf{q}),
\end{equation}
with $p$ and $\rho$ being its pressure and mass density, respectively. The gravitation intensity vector $\textbf{g}(\mathbf{q})$ can be expressed in terms of gravitational potential $\phi(\mathbf{q})$:
\begin{equation}\label{gravedad}
\textbf{g}(\mathbf{q})= - \nabla \phi(\mathbf{q}).
\end{equation}
Considering the total differentiations for $\Phi(\mathbf{q})$ and $p(\mathbf{q})$:
\begin{equation}
d\Phi(\mathbf{q})=\mathbf{\nabla}\Phi(\mathbf{q}) \cdot d\mathbf{q}\mbox{ and
}dp(\mathbf{q})=\mathbf{\nabla}p(\mathbf{q}) \cdot d\mathbf{q},
\end{equation}
the hydrodynamic equilibrium condition (\ref{hidro}) can also be rewritten as:
\begin{equation}\label{NuevaEquiHidro}
\beta \frac{dp\left(\Phi;\gamma\right)}{d\Phi}\equiv n\left(\Phi;\gamma\right),
\end{equation}
where $n\left(\Phi;\gamma\right)$ and $p\left(\Phi;,\gamma\right)$ are the particles density and the pressure expressed in terms of dimensionless potential $\Phi$. Considering differentiation identity (\ref{diff}), it is straightforward verified that hydrodynamic quantities (\ref{densidad particulas}) and (\ref{PresionGamma}) derived from truncated $\gamma$-exponential model satisfy the condition of hydrostatic equilibrium (\ref{hidro}).

\begin{figure}[t]
\begin{center}
\includegraphics[width=4.5in]{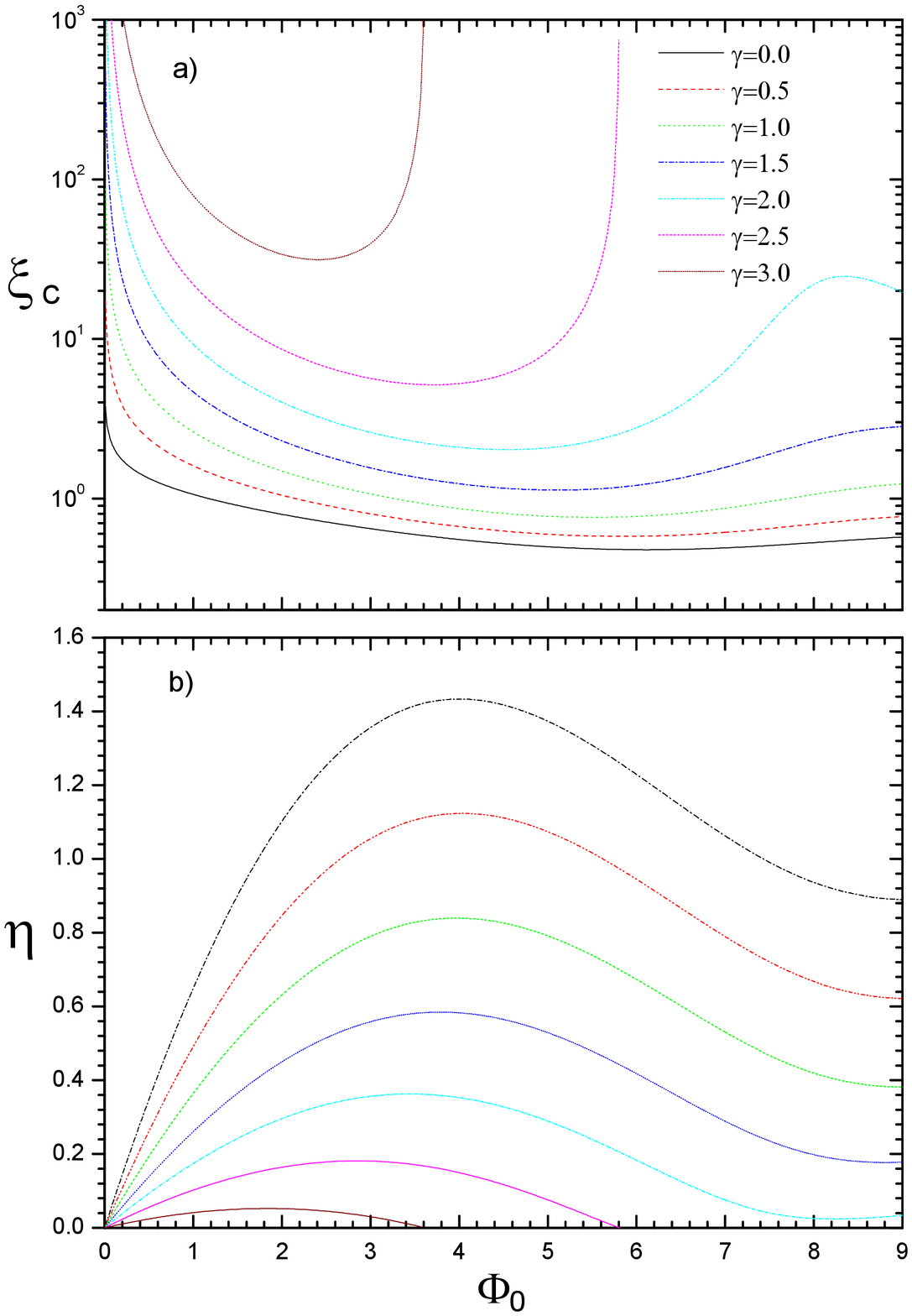}
\caption{Behavior of some numerical parameters $\xi_{c}$ and $\eta$ \emph{versus} the central value of dimensionless potential $\Phi_{0}$ derived from integration of nonlinear Poisson problem (\ref{p1}).}
\label{numericos.eps}
\end{center}
\end{figure}

Mass density $\rho(\mathbf{q})=m n(\mathbf{q})$ and gravitational potential $\phi(\mathbf{q})$ are related via Poisson equation, $\Delta \phi(\mathbf{q})=4\pi G \rho(\mathbf{q})$. Restricting to situations with spherical symmetry, one obtains the following nonlinear Poisson problem in terms of dimensionless potential (\ref{dimensionless.pot}):
\begin{equation}\label{p1}
\frac{1}{\xi^{2}}\frac{d}{d\xi}\left[\xi^{2}\frac{d}{d\xi}\Phi(\xi)\right]=-4 \pi E\left[  \Phi(\xi);\gamma+\frac{3}{2}\right].
\end{equation}
We have considered here the dimensionless radius:
\begin{equation}
\xi=\sqrt{\kappa}r/R_{t},
\end{equation}
where $r$ is the physical radius, $\kappa=Gm^{2}\beta CR^{2}$ is an auxiliary constant, $R_{t}$ is the tidal radius. Problem (\ref{p1}) can be solved in a numerical way by demanding the following conditions at the origin:
\begin{equation}
\Phi\left(  0\right)  =\Phi_{0},~\Phi^{\prime}\left(  0\right) =0,
\end{equation}
which is integrated till the vanishing of dimensionless potential at the surface of the system with dimensionless radius $\xi_{c}=\sqrt{\kappa}$:
\begin{equation}\label{Boundary}
\Phi\left(  \xi_{c}\right)  =0,\Phi^{\prime}\left(  \xi_{c}\right)
=-\frac{\eta}{\xi_{c}}.
\end{equation}
This constant allows to express the mass density $\rho$ in units of the characteristic density $\rho_{c} = M/R^{3}_{t}$ as follows:
\begin{equation}
\rho=\frac{\kappa}{\eta}E\left(  \Phi;\gamma+\frac{3}{2}\right)  ,
\label{den}%
\end{equation}
where $\eta$ is the dimensionless inverse temperature:
\begin{equation}
\eta=\beta\frac{GMm}{R_{t}}.
\end{equation}
\begin{figure}[t]
\begin{center}
\includegraphics[width=4.5in]{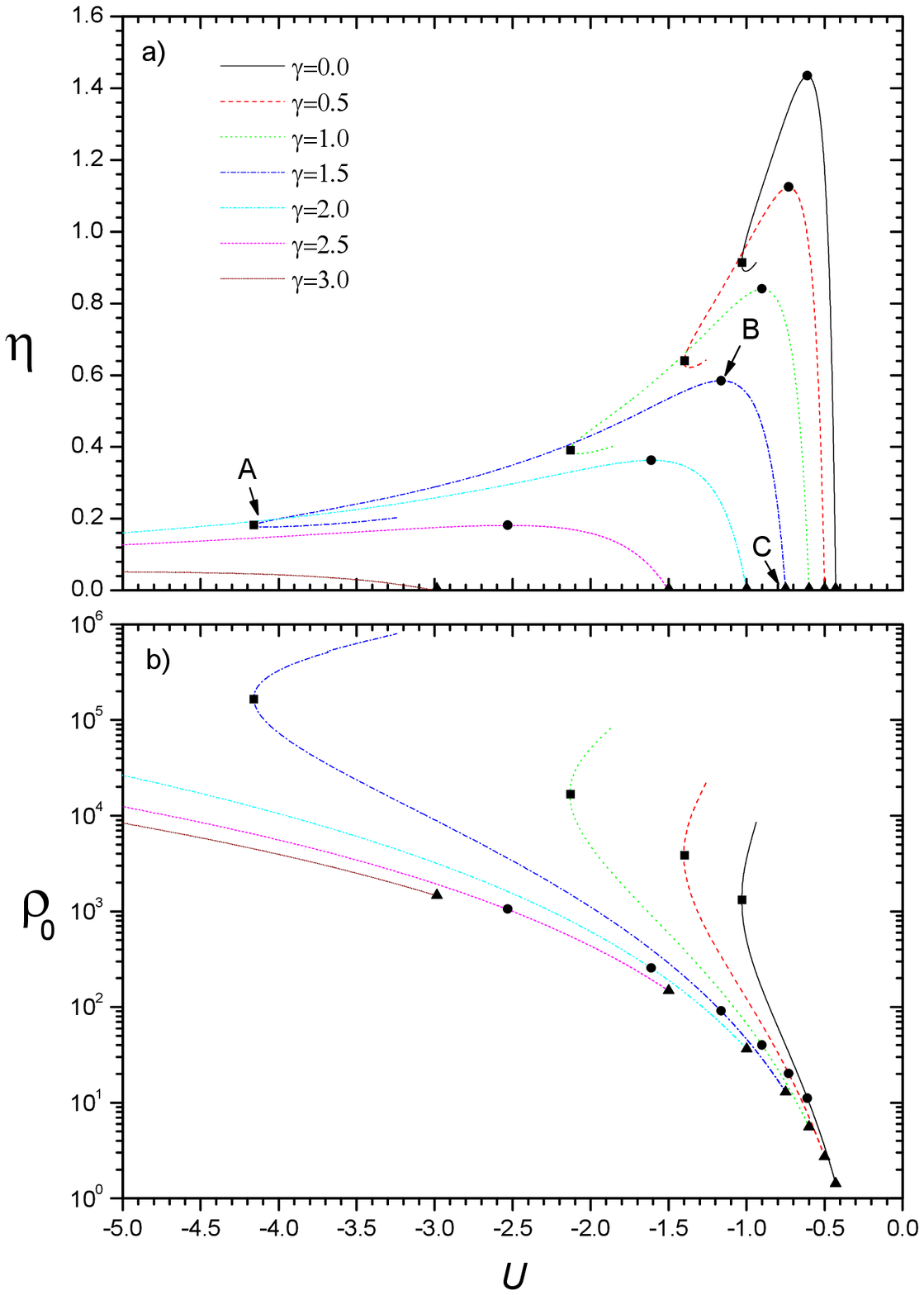}
\caption{Behavior of some thermodynamic quantities for some values of deformation parameter
$\gamma$. Panel a) Caloric curves $\eta$ \textit{versus} $U$; Panel b) Central particles density $\rho_{0}$
\textit{versus} $U$. Additionally, we have included the following notable points: (squares) critical point
of gravothermal collapse $U_{A}$; (circles) critical point of isothermal collapse $U_{B}$; (triangles) critical point of evaporation disruption $U_{C}$.}
\label{Curvacalorica_1.eps}
\end{center}
\end{figure}
The total energy $U=K+V$ can be obtained from the total kinetic $K$ and total potential energy $V$:
\begin{eqnarray}\nonumber
V & = &-\frac{1}{2}-\frac{1}{2\eta^{2}\xi_{c}}\int_{0}^{\xi_{c}}E\left[
\Phi(\xi);\gamma+\frac{3}{2}\right]  \Phi(\xi)~4\pi\xi^{2}d\xi,\\
K & = &\frac{3}{2\eta^{2}\xi_{c}}\int_{0}^{\xi_{c}}E\left[
\Phi(\xi);\gamma+\frac{5}{2}\right]  4\pi\xi^{2}d\xi,\label{K}
\end{eqnarray}
which are expressed in units of the characteristic energy $U_{c}=GM^{2}/R_{t}$.

\section{Results and discussions}

\begin{figure}[t]
\begin{center}
\includegraphics[width=4.5in]{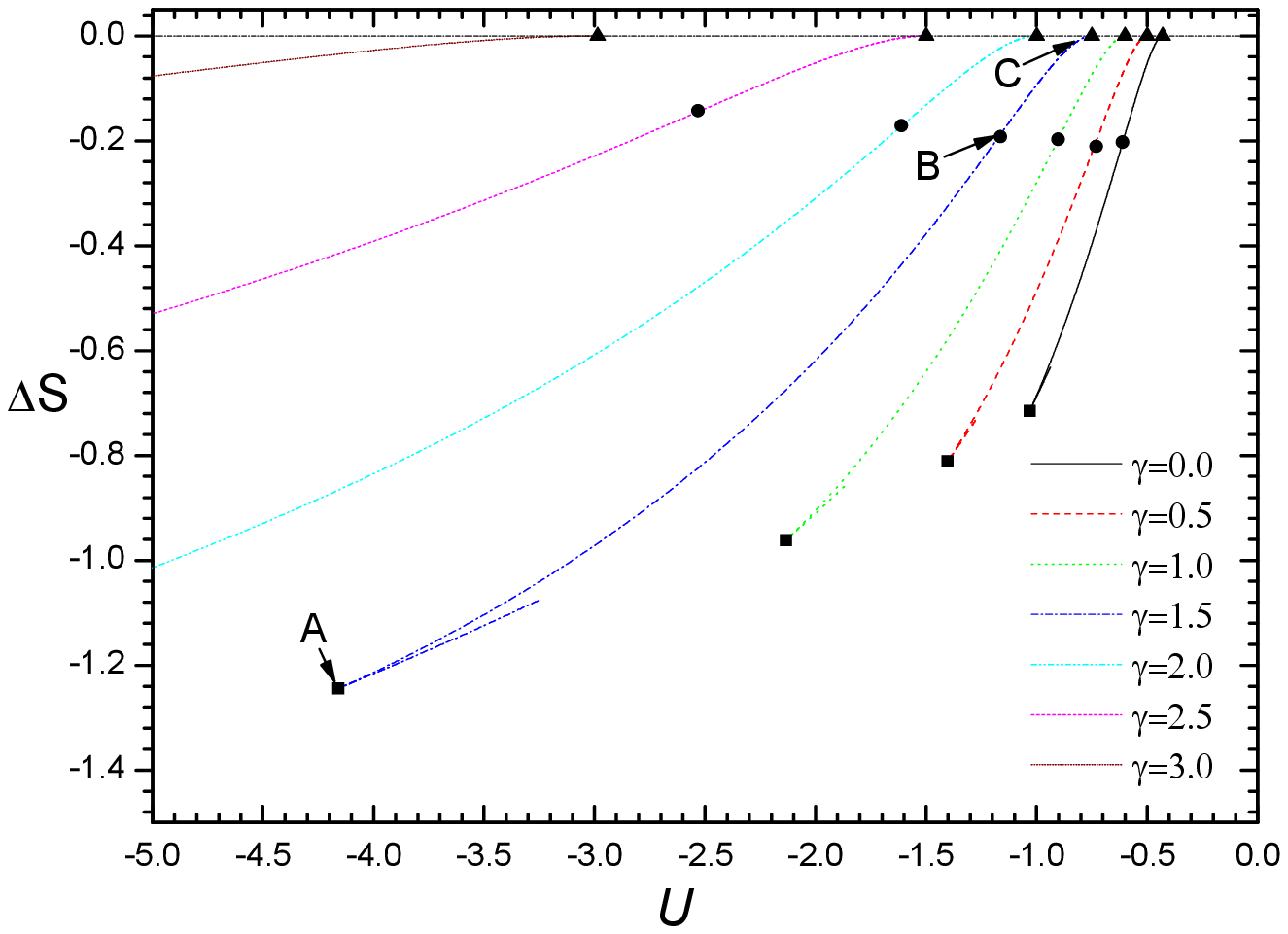}
\caption{Energy dependence of entropy difference $\Delta S=S-S_{C}$ obtained from numerical integration using definition (\ref{entropy.integral}). Only the upper branch of microcanonical caloric curves shown in figure \ref{Curvacalorica_1.eps} correspond to stable quasi-stationary configurations because of they exhibit the higher entropy.}
\label{Entropy_1.eps}
\end{center}
\end{figure}

\subsection{Generalities}

Many studies about astrophysical systems and cosmological problems \cite{Binney,chandra} start from assuming a polytropic dependence between the pressure $p$ and the particles density $\rho$:
\begin{equation}\label{state.eq}
p=C \rho^{\gamma^{*}},
\end{equation}
where $C$ is a certain constant and $\gamma^{*}$ the so-called \emph{polytropic index}. The phenomenological state equation (\ref{state.eq}) can be combined with condition of hydrostatic equilibrium (\ref{hidro}) to derive a power-law relation between the density and the dimensionless potential $\Phi$:
\begin{equation}
\rho=K\Phi^{n},
\end{equation}
where $K=\left[(\gamma^{*}-2)/\gamma^{*} m \beta C\right]^{n}$ and the exponent $n\equiv1/(\gamma^{*}-1)\geq 0$. The marginal case $\gamma^{*}=1$ is fully licit. Condition of hydrostatic equilibrium now predicts an exponential dependence between $\rho$ and $\Phi$:
\begin{equation}
\rho=K^{*}\exp\left[\Phi\right],
\end{equation}
where $K^{*}$ is a certain integration constant. This last behavior can be associated with a system under \emph{isothermal conditions} imposing the constraint $C\equiv 1/m\beta$, where $m$ is the mass of constituting particles and $\beta$ is the inverse temperature. The ansatz (\ref{df.model}) proposed in this study allows to consider the above dependencies into a unifying fashion. According the asymptotic behavior of truncated $\gamma$-exponential function (\ref{asymptotic}), the particles density (\ref{den}) describes a power-law dependence for $\Phi$ small with exponent $n=\gamma+3/2$, while an exponential dependence for $\Phi$ large enough:
\begin{equation}\label{asymp.density}
\rho\left(  \Phi;\gamma\right)  =\left\{
\begin{array}
[c]{cc}%
\propto\Phi^{n}, & x<<1,\\
\propto\exp\left(  \Phi\right)  , & x>>1.
\end{array}
\right.
\end{equation}
Accordingly, the deformation parameter $\gamma$ can be related to the polytropic index $\gamma^{*}$ of polytropic state equation (\ref{state.eq}) as $\gamma^{*}=(2\gamma+5)/(2\gamma+3)$. According to nonlinear Poisson problem (\ref{p1}), the dimensionless potential $\Phi$ decreases from the inner towards the outer region of the system. This means that the truncated $\gamma$-exponential models could describe distribution profiles with \textit{isothermal cores} and \textit{polytropic haloes}. Since the particles density (\ref{den}) should vanish at the system surface, where dimensionless potential $\Phi(\xi_{c})=0$, one should expect that the external dimensionless radius $\xi_{c}$ should be finite, $0<\xi_{c}<+\infty$. However, polytropic profiles with exponent $n>5$ are infinitely extended in the space and exhibit an infinite mass, $\xi_{c}\rightarrow+\infty$ and $M\rightarrow +\infty$. Such profiles cannot describe any realistic situation \cite{chandra}. On the other hand, finite character of quasi-stationary distribution function $\emph{f}_{qe}(\textbf{q},\textbf{p})$ is only possible if deformation parameter $\gamma\geq0$. Consequently, the admissible values of deformation parameter $\gamma$ must be restricted to the following interval:
\begin{equation}
0\leq\gamma<\gamma_{m}=7/2.
\end{equation}
Accordingly, these models can only describe polytropic dependencies with exponent $n$ restricted to the interval $3/2\leq n<5$. As show below, this restriction will be manifested in both thermodynamic quantities and distribution profiles.

\begin{figure}[t]
\begin{center}
\includegraphics[width=4.0in]{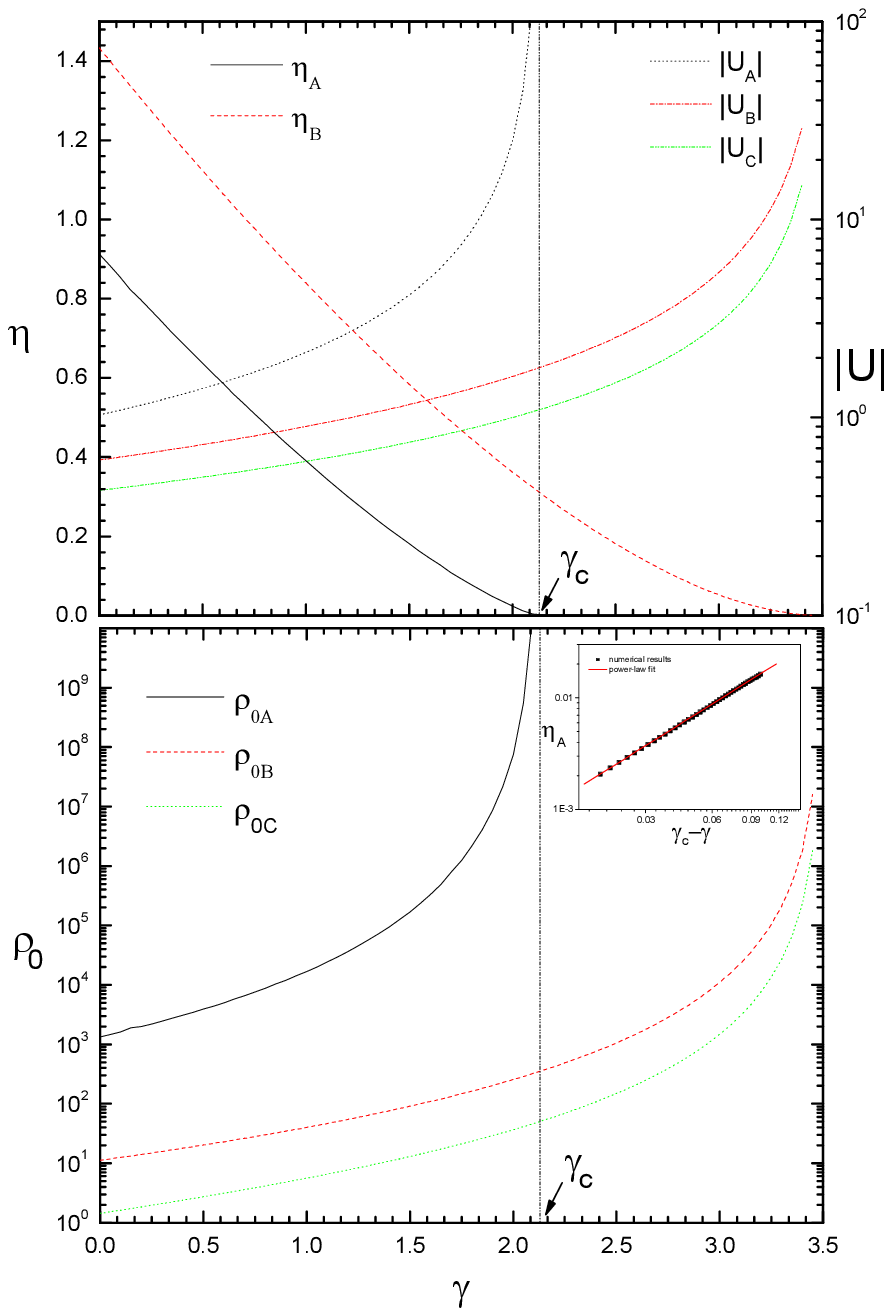}
\caption{Thermodynamic quantities associated with the notable points \textit{versus} deformation parameter $\gamma$: Panel a) the inverse temperatures $\left(\eta_{A},\eta_{B}\right)$ and the modulus of energies   $\left(U_{A},U_{B},U_{C}\right)$; Panel b) the central particles densities $\left(\rho_{0A},\rho_{0B},\rho_{0C}\right)$. Inset panel: Fit of dependence $\eta_{A}(\gamma)$ near the critical value $\gamma_{c}$ using the power-law (\ref{form.power}).}
\label{CalculoFino_1.eps}
\end{center}
\end{figure}

Nonlinear Poisson problem (\ref{p1}) is integrated using $\gamma$ and $\Phi_{0}$ as independent integration parameters. This task was accomplished using Runge-Kutta fourth-order method, which was implemented using FORTRAN 90 programming. Results from this integration are shown in figure \ref{numericos.eps}, in particular, dependencies $\xi_{c}$ and $\eta$ \emph{versus} the central value of dimensionless potential $\Phi_{0}$ for some values of deformation parameter $\gamma$. According to these results, all dependencies of the dimensionless radius $\xi_{c}$ diverge when dimensionless potential $\Phi_{0}$ approaches to zero, which means that admissible values of parameter $\Phi_{0}$ are nonnegative. Curiously, dependencies of dimensionless radius $\xi_{c}$ corresponding to deformation parameter $\gamma=2.5$ and $\gamma=3.0$ diverge at a certain value $\Phi^{\infty}_{0}$ of dimensionless potential $\Phi_{0}$ that depends on deformation parameter $\gamma$, while the corresponding dependencies of dimensionless inverse temperature $\eta$ simultaneously vanishes. This means that value of dimensionless potential $\Phi_{0}$ above the point $\Phi^{\infty}_{0}$ are also nonphysical. A better understanding above the physical meaning of behaviors observed in dependencies of figure \ref{numericos.eps} is achieved analyzing dependencies of thermodynamic quantities.

\begin{figure}[t]
\begin{center}
\includegraphics[width=4.0in]{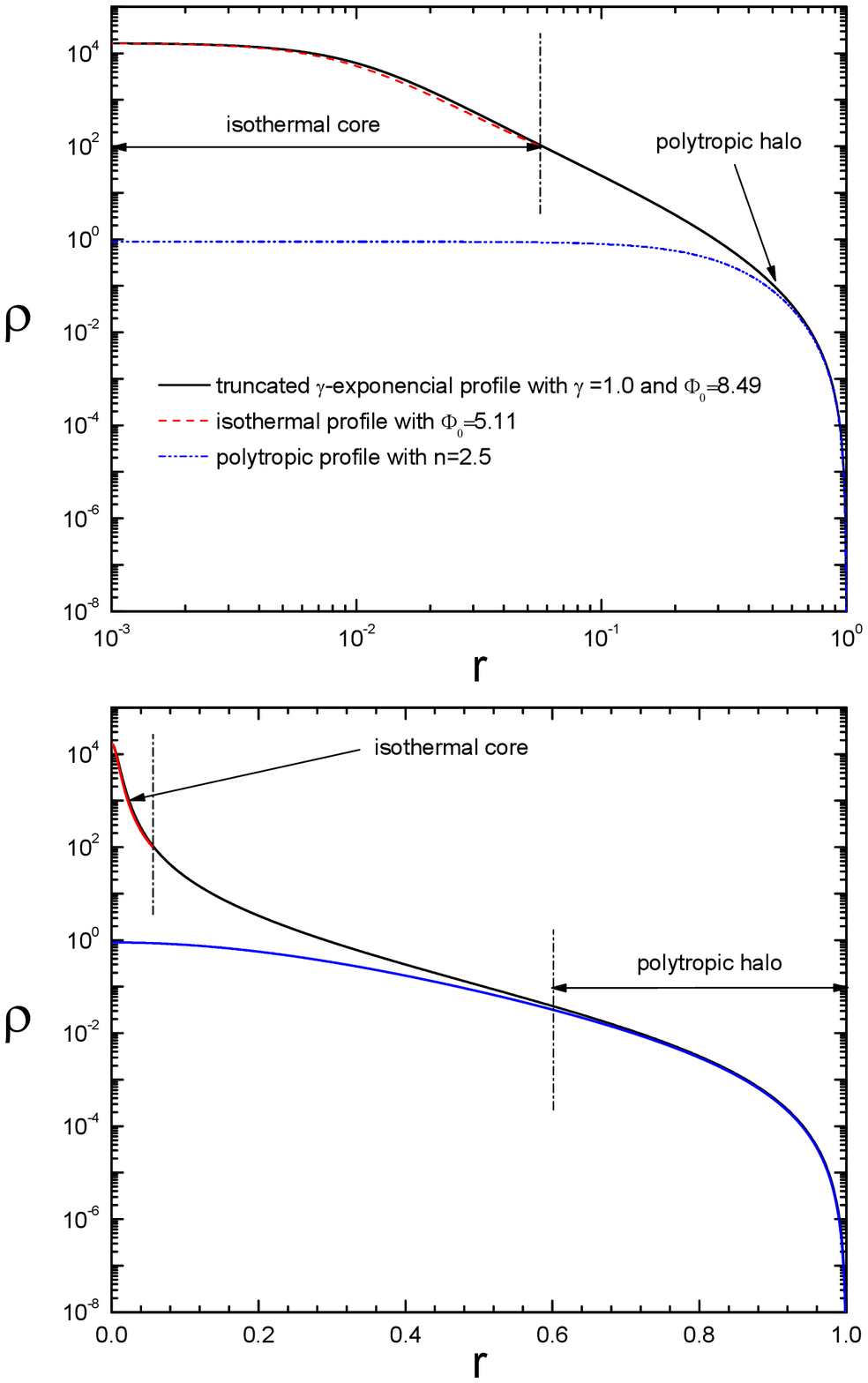}
\caption{Panel a) Truncated $\gamma$-exponential profile with $\gamma=1$ (Michie-King profile) corresponding to the point of gravothermal collapse and its comparison with isothermal and polytropic profiles using \emph{log-log scales}. Panel b) The same dependencies but now using \emph{linear-log scales} to appreciate between the polytropic fit of the halo. Accordingly, the proposed family of models can describe distribution profiles that exhibit isothermal cores and polytropic haloes.}
\label{comparacion_1.eps}
\end{center}
\end{figure}

\subsection{Thermodynamical behavior}

Dependencies of the inverse temperature $\eta$ and central particles densities $\rho(0)$ \emph{versus} the dimensionless energy $U$ are shown in figure \ref{Curvacalorica_1.eps} for different values of deformation parameter $\gamma$. All quasi-stationary configurations obtained from these models have always negative values for the energy $U$. Moreover, one can recognize the existence of three notable points:
\begin{itemize}
\item \textit{Critical point of gravothermal collapse }$U_{A}$: Quasi-stationary configuration with minimum energy $U_{A}$. There not exist quasi-stationary configuration for energies below this point. If a system is initially prepared with an energy below this point, it will experience an instability process that leads to sudden contraction of the system under its own gravitational field, a phenomenon commonly referred to in the literature as \textit{gravothermal collapse} \cite{Lynden-Bell}.

\item \textit{Critical point of isothermal collapse} $U_{B}$: Quasi-stationary configuration with minimum temperature $T_{B}$. There not exist a stable configurations for temperatures $T<T_{B}$. If a system under \textit{isothermal conditions} (in presence of a thermostat at constant temperature) is initially put in thermal contact with a heat reservoir with $T<T_{B}$, this system will experience a instability process fully analogous to gravitational collapse, which is referred to as \emph{isothermal collapse} \cite{Lynden-Bell}. This type of thermodynamical instability is less relevant than the gravothermal collapse because of the presence of a thermostat is actually an unrealistic consideration in most of astrophysical situations.

\item \textit{Critical point of evaporation disruption} $U_{C}$: Quasi-stationary configuration with maximal energy, that is, there not exist quasi-stationary configurations for energies $U>U_{C}$. The existence of this superior bound is a direct consequence of the incidence of evaporation, which imposes a maximum value for the individual mechanical energies of the system constituents, $\varepsilon <\varepsilon_{s}$. If the system is initially prepared prepared with an energy above, it will experience a sudden evaporation in order to release its excess of energy \cite{Vel.QEM2}. Note that the inverse temperature $\eta_{C}$ always vanishes at this point regardless the value of deformation parameter $\gamma$.
\end{itemize}

All quasi-stationary configurations are located inside the energy region $U_{A}\leq U\leq U_{C}$. Moreover, there exist more of one admissible value for the dimensionless inverse temperature $\eta$ for a given energy $U$ near the point of gravothermal collapse $U_{A}$. According to results shown in figure \ref{Entropy_1.eps}, stable quasi-stationary configurations belong to the superior branch $A-B-C$, since these configurations exhibit the higher value of entropy $S$ for a given total energy. Energy dependence of this thermodynamic potential was evaluated from numerical integration of the expression:
\begin{equation}\label{entropy.integral}
\Delta S=S(U)-S_{C}=\int_{U_{C}}^{U}\eta(U') dU',
\end{equation}
which employs as a reference the value $S_{C}$ corresponding to critical point of evaporation instability $U_{C}$. Stable quasi-stationary configurations inside the energy range $U_{A}\leq U\leq U_{B}$ exhibit \textit{negative heat capacities} $C<0$. The existence of this thermodynamic anomaly is a remarkable consequence of the long-range character of gravitation, in particular, because of the short-range divergence of its interaction potential energy:
\begin{equation}
\phi\left(\mathbf{r}_{i},\mathbf{r}_{j}\right)=-Gm_{i}m_{j}/\left|\mathbf{r}_{j}-\mathbf{r}_{i}\right|\rightarrow\infty
\end{equation}
when particles separation distance $\left|\mathbf{r}_{j}-\mathbf{r}_{i}\right|$ drops to zero. While such configurations are unstable if the system is put into thermal contact with an environment at constant temperature (canonical ensemble), they are stable if the system is put into energetic isolation (microcanonical ensemble).

\begin{figure}[t]
\begin{center}
\includegraphics[width=6.8in]{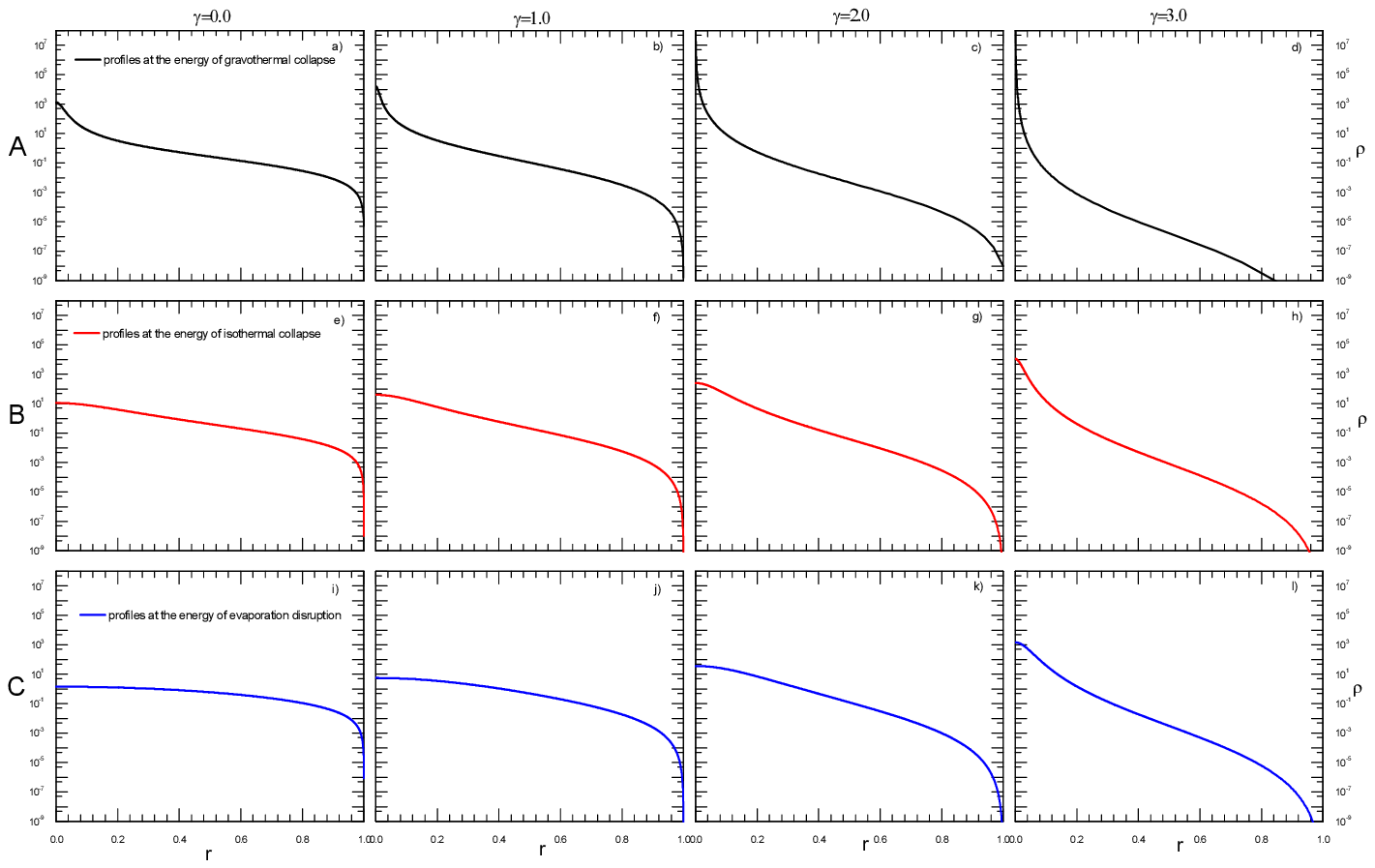}
\caption{Dependence of distribution profiles on the deformation
parameter $\gamma$ for three notable values of the internal energy $U$. While the increasing of deformation parameter $\gamma$ produces profiles with more dense cores and more diluted haloes, the increasing of the energy produces the opposite effect. Interestingly, the qualitative form of the haloes are the same for a given value of deformation parameter $\gamma$. Panels a)-c) Distribution profiles at the energy of gravothermal collapse $U_{A}$ for three values of deformation parameter with $\gamma<\gamma_{c}$. All them exhibit isothermal cores and polytropic haloes. Panel d) A distribution profile at gravothermal collapse with deformation parameter $\gamma>\gamma_{c}$. Note that this profile does not exhibit an isothermal core, but a divergence in the central density. Panels i)-l) Distribution profiles very near the energy of evaporation disruption $U_{C}$ are everywhere polytropic. Panels e)-h) Transitional profiles at the energy of isothermal collapse $U_{B}$. These profiles hardly differ from polytropic profiles i)-l) because of they exhibit more dense cores.} \label{profiles.eps}
\end{center}
\end{figure}

For the sake of a better understanding about the influence of deformation parameter $\gamma$, we have calculated dependence of some thermodynamic quantities at the notable points. Such results are shown in FIG.\ref{CalculoFino_1.eps}. It is clearly evident that the increasing of the deformation parameter $\gamma$ provokes a systematic decreasing of inverse temperatures $(\eta_{A},\eta_{B})$, and the increasing of absolute values of energies $\left(U_{A},U_{B},U_{C}\right)$ and their associated central particles densities $\left(\rho_{0A},\rho_{0B},\rho_{0C}\right)$. Interestingly, the inverse temperature $\eta_{A}$ at the notable point of gravothermal collapse vanishes when $\gamma\geq\gamma_{c}\simeq2.1$. This means that both the total energy $U_{A}$ and the temperature $T_{A}$ diverge at this point as a consequence of divergence of the central density $\rho_{A}$. Precisely, the existence of this divergence also manifested as a divergence of dimensionless radius $\xi_{c}$, which was shown in figure \ref{numericos.eps} for the particular cases with deformation parameters $\gamma=2.5$ and $\gamma=3.0$.

The fact that the energy of gravothermal collapse diverges $U_{A}\rightarrow-\infty$ when $\gamma\geq\gamma_{c}$, significantly reduces the dramatic character of this phenomenon. For values admissible values of deformation parameter $\gamma$ below the point $\gamma_{c}$, the system develops a gravothermal collapse at the finite energy $U_{A}$, which should evolves in a \emph{discontinuous way} towards a certain collapsed structure that is not describable with the present model. For values admissible values of deformation parameter $\gamma$ above the point $\gamma_{c}$, the system should release an infinite amount of energy to reach a collapsed structure with a divergent central density associated with the point of gravothermal collapse. However, this collapsed structure is now described within the present models and the transition is developed in a \emph{continuous way} with the decreasing of the internal energy.

It is noteworthy that an analogous divergence is observed in thermodynamic parameters of other notable points when deformation parameter $\gamma$ approaches its maximum admissible value $\gamma_{m}=3.5$. As expected, this second divergence point is related to the nonphysical character of polytropic dependencies when $n>5$. A more precise estimation for the critical value $\gamma_{c}$ can be obtained considering an adjustment of dependency $\eta_{A}\left(\gamma\right)$ near with a power-law form:
\begin{equation}\label{form.power}
\eta_{A}\left(\gamma\right)= A|\gamma-\gamma_{c}|^{p}.
\end{equation}
As shown in the inset panel of FIG.\ref{CalculoFino_1.eps}, the proposed form (\ref{form.power}) exhibits a great agreement with numerical results for the following set of parameters: $A=0.292\pm0.004$, $p=1$.$24\pm0.01$ and $\gamma_{c}=2.1307\pm0.003$.

\subsection{Distribution profiles}

We show in figure \ref{comparacion_1.eps} a distribution profile with deformation parameter $\gamma=1$ at the point of gravothermal collapse, which correspons to a Michie-King profile with lowest energy. As a consequence of gravitation, the highest concentration of the particles is always located in the inner region of the system, while particles density gradually decays with the increasing of the radius $r$ till vanishes at the tidal radius $R_{t}$. As clearly shown in this figure, the inner region, the core, can be well-fitted with an isothermal profile \cite{Antonov}, while the outer one, the halo, is well-fitted with a polytropic profile \cite{chandra}.

Dependence of distribution profiles on the deformation parameter $\gamma$ and the internal energy $U$ is illustrated in
figure \ref{profiles.eps}, specifically, twelve profiles corresponding to three notable points and forth different values of deformation parameter $\gamma$. Particles concentration in the inner regions decreases with the increasing of the internal energy $U$. The increasing of the deformation parameter $\gamma$ produces distribution profiles with more diluted haloes, and consequently, more dense cores. Not all admissible profiles derived from the present family of models can exhibit isothermal cores. However, all these profiles exhibit polytropic haloes. In fact, distribution profiles near the point of evaporation disruption can be regarded as everywhere polytropic with high accuracy. Finally, distribution profiles corresponding to the point of gravothermal collapse with $\gamma>\gamma_{c}$ are divergent at the origin.

\begin{figure}[t]
\begin{center}
\includegraphics[width=4.0in]{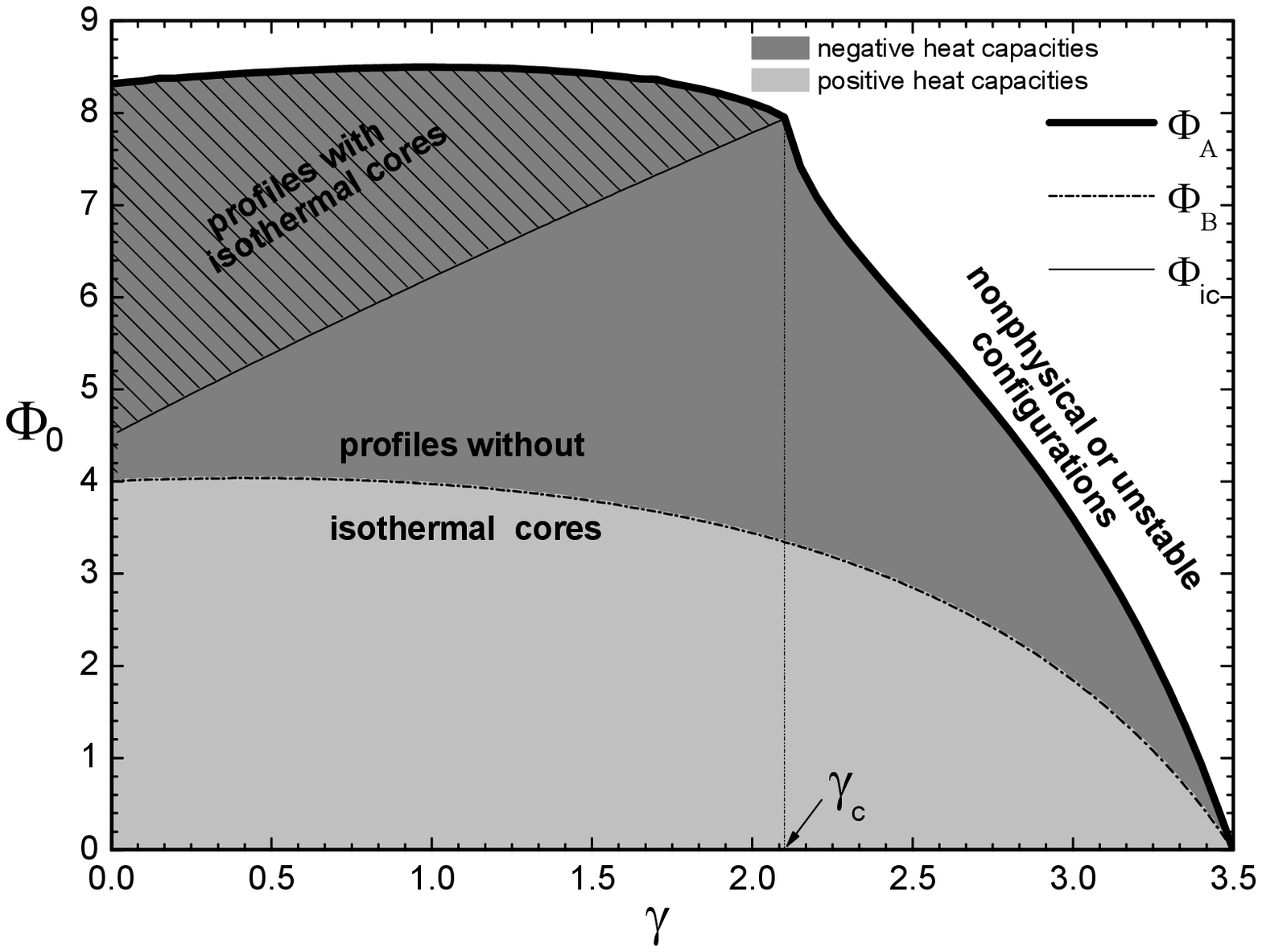}
\caption{Phase diagram of truncated $\gamma$-exponential models in the plane of integration parameters $\gamma-\Phi_{0}$. Dark grey region corresponds to quasi-stationary configurations with negative heat capacities, while light grey region corresponds to positive heat capacities. White regions are nonphysical or unstable configurations. We have emphasized inside the dark grey region those profiles that exhibit isothermal cores and polytropic haloes. Configurations corresponding to isothermal collapse are always outside this region. Quasi-stationary configurations on the line of gravothermal collapse (thick solid line) exhibit a weakly dependence on the deformation parameter $\gamma$ if $\gamma<\gamma_{c}$. This dependence exhibits a significant change for $\gamma_{c}\leq\gamma\leq\gamma_{m}$, which accounts for a divergence in the central density.}
\label{PhiConverge_1.eps}
\end{center}
\end{figure}

We have shown in figure \ref{PhiConverge_1.eps} the phase diagram of truncated $\gamma$-exponential models in the plane of integration parameters $\gamma-\Phi_{0}$ of the nonlinear Poisson problem (\ref{p1}). For each value of deformation parameter $\gamma$, the admissible values of the central dimensionless potential $\Phi_{0}$ are located inside the interval $0\leq\Phi_{0}\leq\Phi_{A}(\gamma)$, where $\Phi_{A}(\gamma)$ corresponds to the critical point of gravothermal collapse $U_{A}$. Central values of dimensionless potential $\Phi_{0}$ above dependence $\Phi_{A}(\gamma)$ correspond to nonphysical on unstable configurations (white region). Additionally, we have included dependence $\Phi_{B}(\gamma)$ associated with the point of isothermal collapse $U_{B}$. Configuration between dependencies $\Phi_{B}(\gamma)\leq\Phi_{0}\leq\Phi_{A}(\gamma)$ exhibit negative heat capacities (dark gray region), while those one with central dimensionless potential $\Phi_{0}$ belonging to interval $0<\Phi_{0}<\Phi_{B}(\gamma)$ exhibit positive heat capacities (light gray region). It is remarkable that dependence
$\Phi_{A}(\gamma)$ is weakly modified by a change in the deformation parameter $\gamma$ for values below the critical point $\gamma_{c}$. However, this function experiences an sudden change above this critical value. As expected, this behavior accounts for a sudden change in behavior of distribution profiles: the proposed models can exhibit profiles with isothermal cores for $\gamma<\gamma_{c}$, while they only exhibit profiles without isothermal cores for $\gamma\geq\gamma_{c}$.

According to equation (\ref{asymp.density}), profiles with isothermal cores are directly related to asymptotic dependence of particles distribution to follow an exponential-law with regard to the local value of dimensionless potential $\Phi(\xi)$. Such an asymptotic behavior is better described in terms of \emph{deviation function} $\delta(x,\gamma)$ with regard to the exponential function, which is introduced in \ref{Exp.Append}. For the sake of convenience, we have denoted dependence $\Phi_{ic}(\gamma)$ as follows:
\begin{equation}
\delta\left[\Phi_{ic}(\gamma);\gamma+\frac{3}{2}\right]=\epsilon,
\end{equation}
where the convergence error $\epsilon$ was fixed at the value $\epsilon=1.6\times 10^{-4}$. This small value guarantees the
matching of this dependence at the critical value of deformation parameter $\gamma_{c}$ with central dimensionless potential
$\Phi_{0}$ associated with the point of gravothermal collapse, $\Phi_{ic}(\gamma_{c})=\Phi_{A}(\gamma_{c})$. Quasi-stationary
configurations located inside the region $\Phi_{ic}(\gamma)\leq\Phi_{0}\leq\Phi_{A}(\gamma)$ and $0<\gamma<\gamma_{c}$ exhibit isothermal cores, that is, the inner regions of particles distribution can be fitted with an isothermal profile.

\section{Summary and open questions}

We have introduced the truncated $\gamma$-exponential models to characterize the properties of astrophysical systems in a quasi-stationary evolution under the incidence evaporation. Our proposal generalizes models of tidal stellar systems available in the literature, such as Michie-King models. These models exhibit many features observed in other astrophysical models. Due to truncation particles distribution in the configuration space $\emph{f}_{qe}(\textbf{q},\textbf{p})$; the distribution of particles in the physical space is located inside a finite region limited by the tidal radio $R_{t}$. Moreover, the total energy $U$ is also restricted to a finite region $U_{A} \leq U \leq U_{C}$. The lower bound $U_{A}$ represents the energy of gravitational collapse, while the upper one $U_{C}$ is the energy of evaporation disruption. Caloric curves of these models exhibit an anomalous branch with negative heat capacities for energies $U_{A} \leq U \leq U_{B}$, where $U_{B}$ is the value of energy corresponding to the point of isothermal collapse, where the system reaches a minimum temperature $T_{B}$. Distribution profiles with lower energy could exhibit isothermal cores and polytropic haloes with exponent $n=\gamma + 3/2$. Moreover, polytropic profiles are always obtained for energies near the point of evaporation disruption $U_{C}$. The admissible values of deformation parameter $\gamma$ are restricted to the interval $0 \leq \gamma < \gamma_{m}=7/2$. This means that this family of models describes polytropic profiles with exponent $n$ inside the interval $3/2\leq n<5$. A nontrivial result obtained from these models is that the existence of distribution profiles with isothermal cores is not longer possible if the deformation parameter $\gamma\geq\gamma_{c}\simeq2.13$. The existence of a notable value $\gamma_{c}$ for deformation parameter indicates a drastic change in the behavior of thermodynamic quantities and distribution profiles. Specifically, these cases exhibit a simultaneous divergence of the energy of gravitational collapse $U_{A}$, its corresponding temperature $T_{A}$ and central density $\rho_{0A}$. For larger energies, these values of deformation parameter $\gamma$ describe distribution profiles almost with a polytropic form.

Before ending this section, let us briefly refer to some open questions. A realistic improvement of the present models is the consideration of mass spectrum for constituting particles, which enable us to study mass segregation influence on the thermodynamic behavior and distribution profiles \cite{Vel.QEM2}. A second improvement is the consideration of factors that lead to a breakdown of spherical symmetry, such as the system rotation or the tidal field of a neighboring astrophysical system \cite{Bertin,Varri,Trenti,Bianchini,Mitchell}. Finally, we shall perform the comparison of results obtained from these models with experimental data, in particular, particles distributions of tidal stellar systems such as globular clusters and elliptical galaxies. These problems will be analyzed in forthcoming works.

\appendix
\section{$\gamma$-exponential function}\label{Exp.Append}

\begin{figure}[t]
\begin{center}
\includegraphics[width=4.0in]{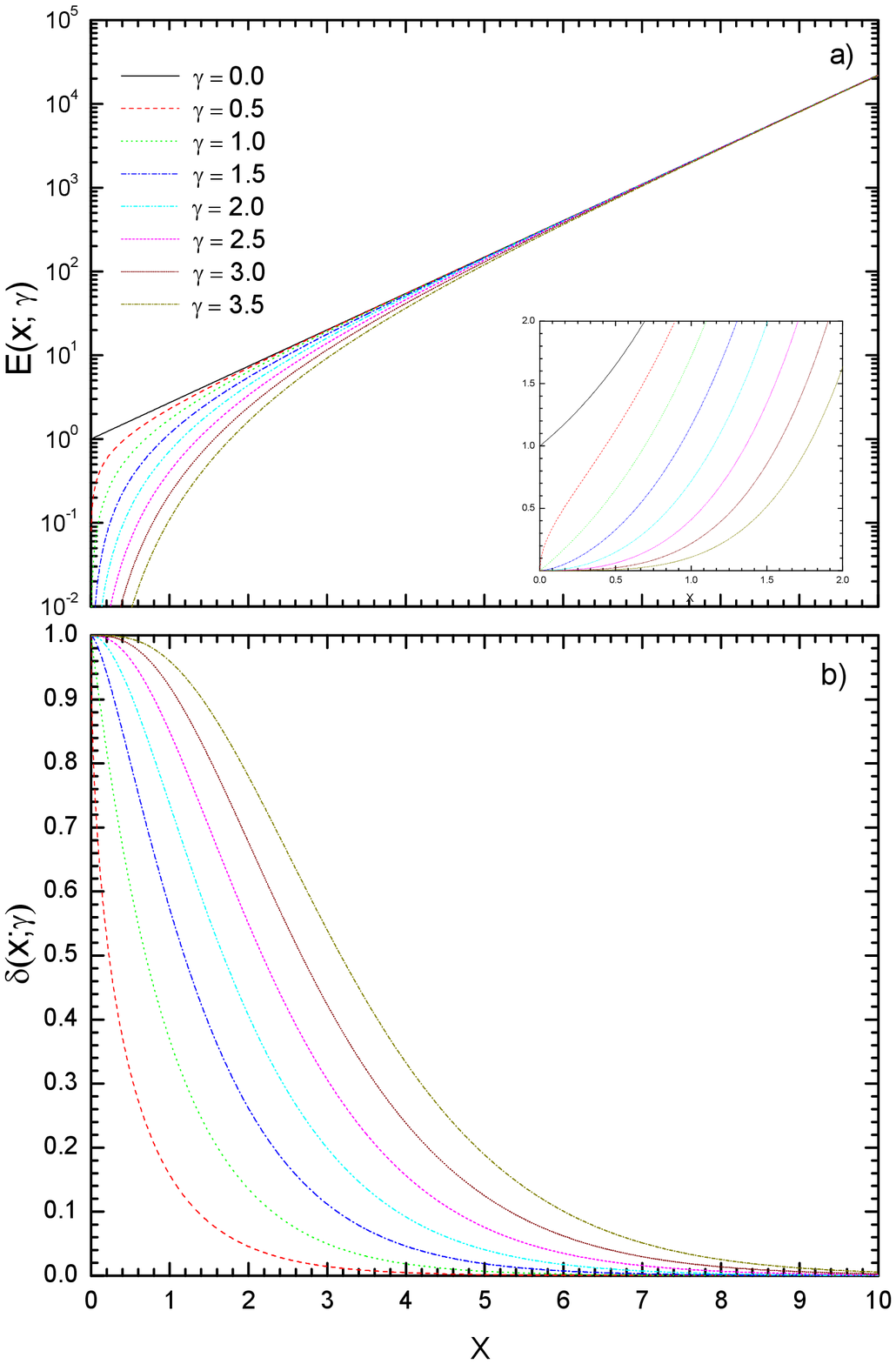}
\caption{Panel a) Some particular examples of the truncated $\gamma $-exponential family in a log-log graph. One can notice the asymptotic dependencies: a power-law behavior for small $x$ and an exponential growing for $x$ large enough. Inset panel: The same dependencies in a linear-linear graph for $x$ near the origin, which allows to appreciated the power-law character of truncation. Panel b) Dependence of deviation function (\ref{defect}) using a linear-linear scale. According to this figure, the relative deviation from exponential decreases with the increasing of the independent variable $x$, while it grows with the increasing of the deformation parameter $\gamma$.}\label{family_1.eps}
\end{center}
\end{figure}

The $\gamma$-exponential function $E\left( x;\gamma\right)$ is defined with the help of called \textit{gamma distribution} as follows:
\begin{equation}
E\left(  x;\gamma\right)  =e^{x}\frac{1}{\Gamma\left(  \gamma\right)  }%
\int_{0}^{x}e^{-\tau}\tau^{\gamma-1}d\tau, \label{def}%
\end{equation}
which vanishes for $x<0$. This family of functions is directly related to \emph{lower incomplete gamma function}:
\begin{equation}\label{gincomplet}
\gamma\left(s,x\right) = \int^{x}_{0} e^{-t}t^{s-1}dt,
\end{equation}%
which only differs from expression (\ref{def}) because of the factor $A\left(s,x\right)= e^{x}/\Gamma\left(s\right)$. Moreover, the same one can also be obtained applying the \emph{fractional integral operator} \cite{Miller}:
\begin{equation}\label{dfractorial}
\left(J^{\gamma}f\right)=\frac{1}{\Gamma\left(\gamma\right)}\int_{0}^{x}\left(x-t\right)^{\gamma-1}f\left(t\right)dt
\end{equation}
over the exponential function. Changing the integration variable $t\rightarrow x-t$ in this last definition, one obtains:
\begin{equation}
\left(J^{\gamma}f\right)\equiv\frac{1}{\Gamma\left(\gamma\right)}\int_{0}^{x}t^{\gamma-1}f\left(x-t\right)dt,
\end{equation}
which drops to expression (\ref{def}) when $f(t)=\exp(t)$.

Performing the integration by parts, one can obtain from definition (\ref{def}) the recurrent relation:
\begin{equation}\label{recurrence}
E\left(  x;\gamma\right)  =\frac{x^{\gamma}}{\Gamma\left(  \gamma+1\right)
}+E\left(  x;\gamma+1\right)  ,
\end{equation}
which leads to the power-expansion in equation (\ref{integral.gamma}). This representation allows to obtain the following particular expressions:
\begin{equation}\label{particular}
E\left(  x;-n\right)  =E\left(  x;0\right)  =e^{x},~E\left(
x,\frac{1}{2}\right)  =e^{x}\mathrm{erf}\left(\sqrt{x}\right) ,
\end{equation}
where $n$ is any positive integer number. In general, the $\gamma$-exponential $E\left(  x;\gamma\right) $ is
a continuous and differentiable function for every $x>0$. This function vanishes at $x=0$ whenever $\gamma>0$,  while it exhibits a discontinuity at $x=0$ when the deformation parameter $\gamma\rightarrow0$. It is easy to verify that the function $E\left( x;\gamma\right) $ with $\gamma>0$ exhibits an \textit{exponential behavior} for $x$ sufficiently large and a \textit{power-law} dependence for small $x$:
\begin{equation}\label{asymptotic}
E\left(  x;\gamma\right)  =\left\{
\begin{array}
[c]{cc}%
x^{\gamma}/\Gamma\left(  \gamma+1\right)  & \mbox{ if }x<<1,\\
e^{x} & \mbox{ if } x>>1.
\end{array}
\right.
\end{equation}
According to recurrence relation (\ref{recurrence}), the $\gamma$-exponential function $E(x,\gamma)$ always diverges at the origin $x=0$ if the deformation parameter $\gamma$ is negative but it is not integer number. The behavior of this family of functions is shown in panel a) of figure \ref{family_1.eps} for some values of deformation parameter $\gamma$. Moreover, we have also considered the \emph{deviation function}:
\begin{equation}\label{defect}
\delta(x,\gamma)=1-\exp(-x)E(x,\gamma)
\end{equation}
that characterizes the relative deviation of $\gamma$-exponential $E(x,\gamma)$ from the ordinary exponential. The behavior of this later function is illustrated in panel b) of figure \ref{family_1.eps}.

Using the power-expansion expression (\ref{integral.gamma}), one can obtain the \textit{integration-differentiation rules}:
\begin{eqnarray}\label{diff}
\frac{d}{dx}E\left(  x;\gamma\right)  & = & E\left(  x;\gamma-1\right),
\\
\int E\left(  x;\gamma\right)  dx & = & E\left(  x;\gamma+1\right)  +C,
\end{eqnarray}
and the \textit{convolution formula}:
\begin{equation}
\frac{1}{\Gamma\left(  \nu\right)  }\int_{0}^{x}E\left(  x-\tau
;\gamma\right)  \tau^{\nu-1}d\tau=E\left(  x;\gamma+\nu\right).
\label{convolution}
\end{equation}
This last relation is obtained using the \emph{Beta function}:
\begin{equation}
B\left(  \mu,\nu\right)  =\int_{0}^{1}\left(  1-\tau\right)  ^{\mu-1}\tau
^{\nu-1}d\tau=\frac{\Gamma\left(  \mu\right)  \Gamma\left(  \nu\right)
}{\Gamma\left(\mu+\nu\right)}.
\end{equation}
Note that definition (\ref{def}) is a particular case of convolution formula (\ref{convolution}) for $\gamma=0$. Moreover, this identity accounts for the fractional integration of $\gamma$-exponential function:
\begin{equation}
\hat{J}^{\nu}E\left(x;\gamma\right)\equiv \frac{1}{\Gamma\left(  \nu\right)  }\int_{0}^{x}E\left( \tau
;\gamma\right)  ( x-\tau)^{\nu-1}d\tau=E\left(  x;\gamma+\nu\right).
\end{equation}

\section*{Acknowledgments}
Velazquez thanks the financial support of CONICyT/Programa Bicentenario de Ciencia y Tecnolog\'{\i}a PSD \textbf{65}. He also thanks partial financial support from VRIDT-UCN research programme.

\end{document}